\documentclass{ifacconf}

\usepackage{graphicx}      % include this line if your document contains figures
\usepackage{natbib}        % required for bibliography
\usepackage{amsmath}
\usepackage{amssymb}
\usepackage{arydshln}
\usepackage{xcolor} 
\usepackage{mathtools}

\newcommand{\tr}{\textrm{tr}}
\newcommand{\E}{\mathcal{E}}
\newcommand{\C}{\mathbb{C}}
\newcommand{\R}{\mathbb{R}}

\begin{document}
\begin{frontmatter}

\title{Modular Model Reduction of Interconnected Systems: A Top-Down Approach \thanksref{footnoteinfo}} 

\thanks[footnoteinfo]{This publication is part of the project Digital Twin with project number P18-03 of the research programme Perspectief which is (mainly) financed by the Dutch Research Council (NWO).}

\author[TU/e]{Lars A.L. Janssen}
\author[RUG]{\ Bart Besselink}
\author[TU/e]{\ Rob H.B. Fey}
\author[TU/e]{\ Nathan van de Wouw}

\address[TU/e]{Dynamics and Control group, Department of Mechanical Engineering,  Eindhoven University of Technology}     % Please supply                                              
\address[RUG]{Bernoulli Institute for Mathematics, Computer Science and Artificial Intelligence, University of Groningen} % full addresses

\begin{abstract}                % Abstract of not more than 250 words.
Models of complex systems often consist of multiple interconnected subsystem/component models that are developed by multi-disciplinary teams of engineers or scientists.
To ensure that such interconnected models can be applied for the purpose of simulation and/or control, a reduced-order model for the interconnected dynamics is needed.
In the scope of this paper, we pursue this goal by subsystem reduction to warrant modularity of the reduction approach. 
Clearly, by reducing the complexity of the subsystem models, not only the accuracy of the subsystem models is affected, but, consequently, also the accuracy of the interconnected model.
It is typically difficult to predict a priori how the interconnected model accuracy is affected precisely by the subsystem reduction.
In this work, we address this challenge by introducing a top-down approach which enables the translation of given accuracy requirements of the interconnected system model to accuracy requirements at subsystem model level, by using mathematical tools from robust performance analysis.
This allows for the independent reduction of subsystem models while guaranteeing the desired accuracy of the interconnected model.
In addition, we show how this top-down approach can be used to significantly reduce the interconnected model in an illustrative structural dynamics case study.
\end{abstract}


\end{frontmatter}
%===============================================================================
\section{Introduction}
Many complex dynamical systems consist of multiple interconnected components.
These components (i.e., subsystems or substructures) are often designed, manufactured, and tested independently before they are integrated into the complete interconnected system.
%Typically, the aim is to create a coherent system that satisfies requirements in terms of the performance of the interconnected system.
To predict, analyze, and control such interconnected systems, high-fidelity subsystem models are created that accurately describe the dynamic behavior of each subsystem.
However, to use these models for the analysis of the dynamic behavior of the interconnected system, often, the complexity of these subsystem models needs to be managed.
Indeed, if the complexity of a dynamical system model is too high for the model to be used in practice, model order reduction (MOR) is required.

In this work, we consider MOR of linear time-invariant (LTI) systems~\citep{antoulas2005,schilders2008,besselink2013}.
Examples of commonly used projection-based methods used for MOR are the proper orthogonal decomposition method~\citep{kerschen2005}, reduced basis methods~\citep{boyaval2010}, balancing methods~\citep{gugercin2004,moore1981,glover1984} and Krylov methods~\citep{grimme1997}.
All of these MOR methods have in common that they aim to compute a reduced-order model (ROM) that still provides an accurate description of the system dynamics but is significantly reduced in complexity in comparison to the high-order model.

In this work, we will deal with managing the complexity of systems of interconnected subsystem models. 
Specifically, the main goal of this work is to construct a model of the interconnected system that 1) satisfies given accuracy requirements and 2) is of a suitable complexity such that it can be used for the application of the model, e.g., for controller design or diagnostics.
The accuracy of the ROM is determined by the difference between the input-to-output behavior of the high-order and the reduced-order interconnected models.

There are several approaches to reduce the complexity of such interconnected models.
Accurate reduced-order models can be obtained with direct reduction of the entire interconnected model as a whole. 
However, this completely destroys the interconnection structure~\citep{lutowska2012}.
To avoid this problem, there are several structure-preserving reduction methods available for interconnected systems~\citep{sandberg2009,vandendorpe2008}
Unfortunately, these methods still require knowledge of the entire interconnected system when computing reduced-order subsystem models.

Since the subsystem models are developed individually and, often, in parallel, we aim to reduce the complexity of the subsystem \emph{modularly}, i.e., on an individual basis.
Such a modular approach has the additional advantages that the computational cost of computing the reduced-order model is significantly reduced \citep{vaz1990} and different reduction methods can be applied for each subsystem individually \citep{reis2008}.
In the structural dynamics field, component mode synthesis (CMS) methods are also modular \citep{klerk2008}.

However, with modular MOR of interconnected systems, we reduce the complexity of subsystem models, which generally leads to an error of the subsystem ROM in comparison to the high-order subsystem model.
If the reduced-order subsystem models are then interconnected, these errors will propagate to the reduced-order interconnected system model, potentially exceeding the given accuracy requirements on the interconnected system model.
Therefore, the need arises for methods to relate interconnected model accuracy requirements to accuracy requirements on the level of individual subsystem models.

The main contribution of this work is a top-down approach that allows us to translate frequency-dependent accuracy requirements on the interconnected model to frequency-dependent accuracy requirements on the input-to-output behavior at a subsystem level.
Then, if the subsystem models are reduced (individually) using any reduction method that can satisfy these subsystem accuracy requirements, the accuracy requirements on the interconnected model are also guaranteed to be satisfied.
We use methods from robust performance analysis to establish this relation.

In \cite{janssen2022_automatica} we have established the mathematical foundation of this method, including stability guarantees and a top-down approach that allows for the computation of accuracy requirement on \emph{one of the subsystem models} based on requirements on the \emph{largest singular value error of the interconnected model}.
In the current paper, we extend this approach in two ways.
Namely, we now allow for the computation of accuracy requirements for 
\begin{enumerate}
\item \emph{all subsystems simultaneously}, and for
\item \emph{all input-output pairs} for each of these subsystems, 
\end{enumerate}
and the subsequent \emph{modular} reduction of all subsystems by solving a single optimization problem.
Furthermore, we show on the illustrative example as used in \cite{janssen2022_automatica} how these extensions can be used to significantly reduce the system using the top-down, modular MOR approach.
Note that in \cite{janssen2022}, we have shown with a preliminary version of the relation established in \cite{janssen2022_automatica} that this relation can be used to determine a priori error bounds to the error of the reduced-order interconnected model based on error bounds of ROMs of the subsystems (following a bottom-up approach, i.e., considering the inverse problem of translating the accuracy of subsystem models to that of the interconnected system).

The paper is organized as follows. Section~\ref{sec:Problem} gives the problem statement including the modelling framework. 
In Section~\ref{sec:Method}, we show how the problem can be reformulated into a robust performance problem and consequently how it can be solved.
The top-down approach is demonstrated on an illustrative structural dynamics example system in Section~\ref{sec:Example}.
Finally, the conclusions are given in Section~\ref{sec:Conclusion}.

\emph{Notation}. The set of real numbers is denoted by $\R$, of positive real numbers by $\R_{>0}$, and of complex numbers by $\C$. 
Given a transfer function (matrix) $G(s)$, where $s$ is the Laplace variable, $\|G\|_\infty$ denotes its $\mathcal{H}_\infty$-norm. 
The real rational subspace of $\mathcal{H}_{\infty}$ is denoted by $\mathcal{RH}_{\infty}$. 
Given a complex matrix $A$, $A^H$ denotes its conjugate transpose, $\bar{\sigma}(A)$ denotes its largest singular value, $\rho(A)$ denotes its spectral radius, $A = \text{diag}(A_1,A_2)$ denotes a block-diagonal matrix with submatrices $A_1$ and $A_2$, and $A \succ 0$ means that $A$ is positive definite. 
%Given real matrices $B$ and $C$, $B \geq C$ means that $B-C$ is non-negative, i.e., all the elements are equal to or greater than zero.
The identity matrix of size $n$ is denoted by $I_n$.

\section{Problem statement}\label{sec:Problem}
\begin{figure}
   \centering
   \includegraphics[scale=1, page=1]{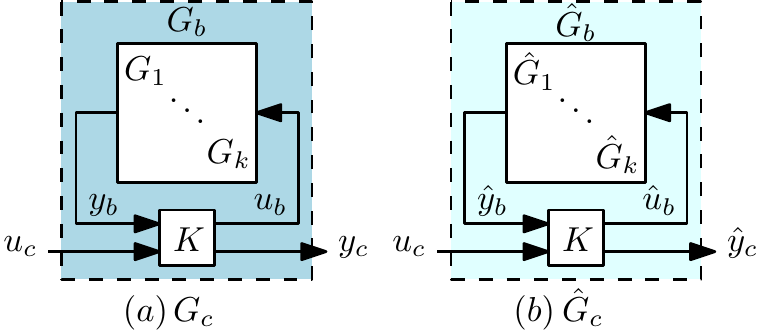} 
   \caption{Block diagram representation of (a) high-order interconnected system $G_c(s)$ and (b) reduced-order interconnected system $\hat{G}_c(s)$.  K represents a static interconnection block.}
   \label{fig:Gc}
\end{figure}
Consider $k$ high-order, linear subsystems $j \in \{1,\dots,k\}$ with transfer functions $G_j(s)$, inputs $u_j$ and outputs $y_j$ of dimensions $m_j$ and $p_j$, respectively, and McMillan degree $n_j$.
We collect the subsystem transfer functions in the block-diagonal transfer function
\begin{equation}
G_b(s) := \textrm{diag}\left(G_1(s),\dots,G_k(s)\right),
\end{equation}
for which the total number of inputs and outputs are then given by $m_b := \sum_{j=1}^{k}m_j$ and $p_b := \sum_{j=1}^{k}p_j$, respectively.
We define inputs $u_b^\top := \left[u_1^\top,\dots,u_k^\top \right]$ and outputs $y_b^\top := \left[y_1^\top,\dots,y_k^\top \right]$.

In this paper, we compute the ROM of the system modularly, i.e., we reduce each subsystem model independently. 
Therefore, consider reduced-order subsystems $j \in \{1,\dots,k\}$ and their transfer functions $\hat{G}_j(s)$, each with inputs $\hat{u}_j$ and outputs $\hat{y}_j$ with dimensions $m_j$ and $p_j$, respectively, and McMillan degree $r_j$. 
Let the reduced-order block-diagonal transfer function be given as 
\begin{equation}
\hat{G}_b(s) := \textrm{diag}\left(\hat{G}_1(s),\dots,\hat{G}_k(s)\right). 
\end{equation}
Then, we define inputs $\hat{u}_b^\top := \left[\hat{u}_1^\top,\dots,\hat{u}_k^\top \right]$ and outputs $\hat{y}_b^\top := \left[\hat{y}_1^\top,\dots,\hat{y}_k^\top \right]$ with dimensions $m_b$ and $p_b$, respectively. 
Both the high-order and reduced-order subsystem models are interconnected according to
\begin{equation}
\label{eq:connection}
\left[\begin{array}{c}
u_b \\ y_c 
\end{array}\right] = 
K \left[\begin{array}{c}
y_b \\ u_c 
\end{array}\right], \left[\begin{array}{c}
\hat{u}_b \\ \hat{y}_c 
\end{array}\right] = 
K \left[\begin{array}{c}
\hat{y}_b \\ u_c 
\end{array}\right], K = \left[\begin{array}{cc}
K_{11} & K_{12} \\
K_{21} & K_{22}
\end{array}\right].
\end{equation}
Here, we have also introduced external inputs $u_c$, high-order external outputs $y_c$ and reduced-order external outputs $\hat{y}_c$.
The number of external inputs and outputs is given by $m_c$ and $p_c$, respectively.
Then, the transfer function from $u_c$ to $y_c$ is given by the upper linear fractional transformation (LFT) of $G_b(s)$ and $K$, which yields
\begin{equation}
\label{eq:Gc}
G_c(s) := K_{21}G_b(s)( I - K_{11}G_b(s))^{-1}K_{12} + K_{22}.
\end{equation}
Since we only reduce the subsystem models, the interconnection structure is preserved.
Therefore, the reduced-order interconnected system transfer function from $u_c$ to $\hat{y}_c$ is, similar to (\ref{eq:Gc}), given by
\begin{equation}
\label{eq:hatGc}
\hat{G}_c(s) := K_{21}\hat{G}_b(s)( I - K_{11}\hat{G}_b(s))^{-1}K_{12} + K_{22}.
\end{equation}
This model framework is illustrated in Figure~\ref{fig:Gc}.

The approach developed in this paper is completely frequency-dependent.
Therefore, we analyze the transfer functions for $s = i\omega$ for some $\omega\in\R$.
Furthermore, we assume that we can define frequency-dependent requirements on the reduction error dynamics 
\begin{equation}
E_c(i\omega) := \hat{G}_c(i\omega) - G_c(i\omega).
\end{equation}
Specifically, we consider the requirement that $E_c(i\omega)$ is contained in the set
\begin{align}
\label{eq:Ec_bound}
\E_c(\omega) := \big\{ E_c(i\omega) \ \big| \ \bar{\sigma}\bigl(V_c(\omega)E_c(i\omega)W_c(\omega) \bigr) < 1 \big\},
\end{align}
where diagonal scaling matrices $V_c(\omega) \in \R^{p_c \times {p_c}}_{>0}$ and $W_c(\omega) \in \R^{m_c \times {m_c}}_{>0}$ can be used to scale the input-output pairs of $E_c(i\omega)$ to fit the requirements.

\begin{figure}
   \centering
   \includegraphics[scale=1, page=8]{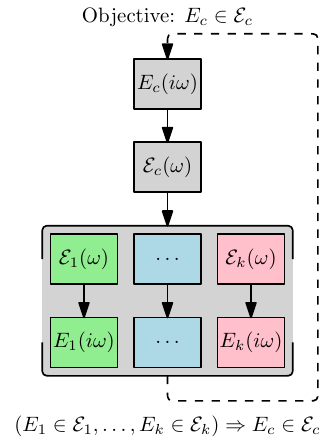} 
   \caption{The top-down approach for modular model reduction of interconnected systems: the determination of reduced-order \emph{subsystem} error requirements $\E_j(\omega)$. When these are satisfied, i.e., $E_j(i\omega) \in \E_j(\omega)$ for all subsystems $j\in\{1,\dots,k\}$, satisfaction of reduced-order \emph{interconnected system} requirements, i.e., $E_c \in \E_c(\omega)$ for $\omega \in \R$, is guaranteed.}
   \label{fig:TD}
\end{figure}
The main goal of this work is to find specifications to the subsystem reduction error dynamics 
\begin{equation}
E_j(i\omega) := \hat{G}_j(i\omega) - G_j(i\omega)
\end{equation}
for each subsystem $j\in\{1,\dots,k\}$ based on $\mathcal{E}_c(\omega)$, i.e., using a top-down approach.
Specifically, we aim to find some sets
\begin{align}
\label{eq:Ej_bound}
\E_j(\omega) := \big\{ E_j(i\omega) \ \big| \ \bar{\sigma}\bigl(W_j^{-1}(\omega)E_j(i\omega)V_j^{-1}(\omega) \bigr) \leq 1 \big\}
\end{align}
such that $E_j(i\omega) \in \mathcal{E}_j(\omega)$ for all $j\in\{1,\ldots,k\}$ implies $E_c(i\omega) \in \mathcal{E}_c(i\omega)$. In (\ref{eq:Ej_bound}), $V_j(\omega)$ and $W_j(\omega)$ are diagonal scaling matrices.

Note that both for the interconnected system in (\ref{eq:Ec_bound}) and the subsystems in (\ref{eq:Ej_bound}), the error requirements implicitly provide a bound on each input-output pair individually.
Namely, for any matrix $A \in \C^{m \times n}$, $\bar{\sigma}(A) < 1$ is only satisfied if the magnitude of all elements in $A$ are less than one. 
These bounds are scaled individually by the elements in the scaling matrices $V_c$, $W_c$, $V_j$ and $W_j$.

%In this work, we will show that we can compute requirements $(\E_1(\omega),\dots,\E_k(\omega))$ for which we can guarantee that the given requirements $\E_c(\omega)$ on the interconnected model are satisfied, i.e., $E_c(i\omega) \in \E_c(\omega)$ if $E_j(i\omega) \in \E_j(\omega)$ for all $j\in\{1,\dots,k\}$.

Once the sets $\E_j(\omega)$ have been determined, it becomes possible to compute reduced-order subsystem models independently.
Namely, if each subsystem $j \in \{1,\dots,k\}$ is reduced such that it satisfies $E_j(i\omega) \in \E_j(\omega)$, it is guaranteed that the reduction error dynamics of the interconnected system satisfy $E_c(i\omega) \in \E_c(\omega)$.
We show this systematic approach schematically in Figure~\ref{fig:TD}.

\section{Methodology}\label{sec:Method}
\begin{figure}
   \centering
   \includegraphics[scale=1, page=2]{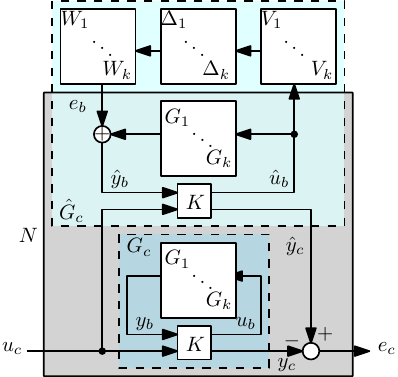} 
   \caption{Block diagram representation of the error dynamics of the interconnected system, $E_c = \hat{G}_c - G_c$, as a function of $V_{b}$, $W_{b}$, $\Delta$ and the nominal system $N$.}
   \label{fig:EcN}
\end{figure}

To find the subsystem accuracy specifications characterized by $(\E_1(\omega),\dots,\E_k(\omega))$ based on the requirement $\E_c(\omega)$ for $\omega \in \R$, we reformulate the problem as a robust performance problem, following along the lines of our work in \cite{janssen2022_automatica}.
First, we define weighting functions $V_{j}(\omega) \in \R^{m_j \times {m_j}}_{>0}$ and $W_{j}(\omega) \in \R^{p_j \times {p_j}}_{>0}$ such that $E_j(i\omega)$ can be written as
\begin{equation}
E_j(i\omega) = W_j(\omega) \Delta_j(i\omega) V_j(\omega),
\end{equation}
for some $\Delta_j(s) \in \mathcal{RH}_\infty$ satisfying $\|\Delta_j\|_\infty \leq 1$.
Then, we rewrite the reduced-order subsystem as
\begin{equation}
\label{eq:EjasDeltaj}
\hat{G}_j(i\omega) = G_j(i\omega) + W_{j}(\omega)\Delta_j(i\omega)V_{j}(\omega).
\end{equation}
By replacing $\hat{G}_j(i\omega)$ with $G_j(i\omega) + W_{j}(\omega)\Delta_j(i\omega)V_{j}(\omega)$ for all $j\in\{1,\dots,k\}$ in Figure~\ref{fig:Gc}(b) and comparing it with the high-order system $G_c(i\omega)$ in Figure~\ref{fig:Gc}(a), we obtain the block diagram in Figure~\ref{fig:EcN}. 
Additionally, we define the nominal transfer function $N(s)$, i.e., the grey block in~Figure~\ref{fig:EcN}, which is given by
\begin{equation}
\label{eq:N}
N(s)=\left[\begin{array}{cc}
N_{11}(s)&N_{12}(s)\\ 
N_{21}(s)&0
\end{array}\right],
\end{equation}
where
\begin{align*}
N_{11}(s) &= K_{11}(I-G_b(s)K_{11})^{-1},\\ 
N_{12}(s) &= (I-K_{11}G_b(s))^{-1} K_{12},\textrm{ and}\\
N_{21}(s) &= K_{21}(I-G_b(s)K_{11})^{-1}.
\end{align*}
The interconnected system error dynamics $E_c$, as shown in Figure~\ref{fig:EcN}, for $\omega \in \R$, can then be given by
\begin{align}
\label{eq:EcN}
E_c(i\omega) = N_{21}&(i\omega)W_{b}(\omega)\Delta_b(i\omega) V_{b}(\omega)\big(I \\
 &-N_{11}(i\omega)W_{b}(\omega)\Delta_b(i\omega)V_{b}(\omega)\big)^{-1}N_{12}(i\omega), \nonumber
\end{align}
where $\Delta_b := \textrm{diag}\left(\Delta_1,\ldots,\Delta_k \right)$, $V_{b} := \textrm{diag}\left(V_{1},\dots,V_{k}\right)$, and $W_{b} := \textrm{diag}\left(W_{1},\dots,W_{k}\right)$.
%\left[v_1^\top,\dots,v_k^\top,v_c^\top\right]I_{m_b+p_c}  \Big| \\
%& \qquad \qquad v_j \in \R^{m_j}_{>0} \ \forall \ j\in\{1,\dots,k\}, v_c \in \R^{p_c}_{>0} \Big\}, \nonumber \\
%Since the proposed approach is completely frequency-dependent, from this point, we analyze the error dynamics error dynamics (\ref{eq:EcN}) for specific frequencies $\omega \in \R$, and thus $s = i\omega$.

To relate the requirements $\E_c$ on the accuracy of the reduced-order interconnected model to accuracy requirements $(\E_1,\dots,\E_k)$ on the reduced-order subsystem models, consider the following sets of matrices.
\begin{align}
\label{eq:setV}
\mathbf{V} &:= \Big\{\textrm{diag}\left(V_1,\dots,V_k,V_c\right) \ \Big| \nonumber\\
& \qquad \quad V_j = \textrm{diag}(v_j), v_j \in \R^{m_j}_{>0} \ \forall \ j \in \{1,\dots,k\}, \nonumber\\
& \qquad \quad V_c = \textrm{diag}(v_c), v_c \in \R^{p_c}_{>0} \Big\}, \\
\label{eq:setW}
\mathbf{W} &:= \Big\{\textrm{diag}\left(W_1,\dots,W_k,W_c\right) \ \Big| \nonumber\\
& \qquad \quad W_j = \textrm{diag}(w_j), w_j \in \R^{p_j}_{>0} \ \forall \ j \in \{1,\dots,k\}, \nonumber\\
& \qquad \quad W_c = \textrm{diag}(w_c), w_c \in \R^{m_c}_{>0} \Big\}, \\
\label{eq:setD}
\mathbf{D} &:= \Big\{ (D_\ell, D_r) \ \Big| \ d_1,\dots,d_k, d_c \in \R_{>0}, \nonumber \\
& \qquad \quad D_\ell = \textrm{diag}\left( d_1 I_{p_1},\dots,d_{k} I_{p_k}, d_cI_{m_c} \right), \nonumber \\
& \qquad \quad D_r = \textrm{diag}\left( d_1 I_{m_1},\dots,d_{k} I_{m_k}, d_cI_{p_c} \right)\Big\}.
\end{align}
Then, we can pose the following theorem.
\begin{thm}
\label{thm:main}
Let $\omega \in \R$, $V(\omega) \in \mathbf{V}$, and $W(\omega) \in \mathbf{W}$, with $\mathbf{V}$ and $\mathbf{W}$ as in (\ref{eq:setV}) and (\ref{eq:setW}), respectively. Consider the system (\ref{eq:Gc}), error dynamics (\ref{eq:EcN}) as in Figure~\ref{fig:EcN}, and requirements $\E_c(\omega)$ as in (\ref{eq:Ec_bound}) and $\E_j(\omega)$, $j\in\{1,\dots,k\}$, as in (\ref{eq:Ej_bound}). 
If there exists a $(D_\ell, D_r) \in \mathbf{D}$, with $\mathbf{D}$ as in (\ref{eq:setD}), such that
\begin{align}
\label{eq:mainLMI}
\left[\begin{array}{cc}
W^{-2}(\omega)D_r^{-1} 	& 	N^H(i\omega) \\
N(i\omega)				& 	V^{-2}(\omega)D_\ell
\end{array}\right] \succ 0,
\end{align}
with N as in (\ref{eq:N}), then, it holds that if all reduced-order subsystem models satisfy their respective error requirements, i.e., $E_j(i\omega) \in \E_j(\omega)$ for all $j\in\{1,\dots,k\}$, then the reduced-order interconnected model will satisfy the interconnected system requirements, i.e., $E_c(i\omega) \in \E_c(\omega)$.
\end{thm}
\begin{pf}
We prove the theorem with the following remarks.
\begin{enumerate}
\item We have proven in~\cite{janssen2022_automatica}, Theorem 3.6, that any $E_j(i\omega)$ satisfying $E_j(i\omega) \in \E_j(\omega)$ for all $j\in\{1,\dots,k\}$, it holds that $E_c(i\omega) \in \E_c(\omega)$, which is equivalent to 
\begin{equation}
\label{eq:mu<1}
\mu_{\mathbf{\Delta}}\bigl(V(\omega)N(i\omega)W(\omega)\bigr) < 1. 
\end{equation}
Here, $\mu_{\mathbf{\Delta}}$ denotes the structured singular value \citep[Definition 3.1]{packard1993} defined by
\begin{align*}
\mu_{\mathbf{\Delta}}(M) := \frac{1}{\min\left\lbrace \bar{\sigma}(\Delta) \ \middle| \ \det(I-M\Delta)=0, \Delta \in \mathbf{\Delta} \right\rbrace}
\end{align*}
for $M \in \C^{(m_b+p_c) \times (p_b+m_c)}$ and $\mathbf{\Delta}$ given by 
\begin{align}
\label{eq:setDelta}
\mathbf{\Delta} &:= \Big\{\text{diag}\big(\Delta_1,\dots,\Delta_k,\Delta_c \big) \ \Big| \ \Delta_c\in\C^{m_c \times p_c}, \nonumber\\
& \qquad \quad \Delta_j\in\C^{p_j\times m_j}, j\in\{1,\ldots,k\} \Big\}.
\end{align}
\item As we have proven in~\cite{janssen2022_automatica}, Theorem 3.6, for any $M \in \C^{(m_b+p_c) \times (p_b+m_c)}$, if there exists a $(D_\ell, D_r) \in \mathbf{D}$ such that $M  D_rM^H \prec D_\ell$, then, given $\Delta \in \mathbf{\Delta}$, $\mu_{\mathbf{\Delta}}(M) < 1$.
\item Since $V(\omega)N(i\omega)W(\omega) \in \C^{(m_b+p_c) \times (p_b+m_c)}$, we have that (\ref{eq:mu<1}) is satisfied if 
\begin{align}
\label{eq:LMI1}
\hspace{-.4cm}V(\omega)N(i\omega)W(\omega)D_rW^H(\omega)N^H(i\omega)V^H(\omega)\prec D_\ell. 
\end{align}
\item Since $W(\omega) \in \mathbf{W}$ and $(D_\ell, D_r) \in \mathbf{D}$ are diagonal, real, and positive definite, $W(\omega)D_rW^H(\omega) = W^2(\omega)D_r \succ 0$.
\item Since $V(\omega) \in \mathbf{V}$ is diagonal, real, and positive definite, after pre- and post-multiplying both sides of the inequality (\ref{eq:LMI1}) by $V^{-1}(\omega)$, we obtain
\begin{equation}
\label{eq:schur}
N(i\omega)W^{-2}(\omega)D_rN^H(i\omega) \prec V^{-2}(\omega)D_\ell.
\end{equation}
\item We obtain (\ref{eq:mainLMI}) as the equivalent Schur's complement of (\ref{eq:schur}).  \hfill \ \qed 
\end{enumerate}
\end{pf}
%\section{Top-down approach}\label{sec:TD}
With Theorem~\ref{thm:main}, for $\omega \in \R$ any combination of requirements on $E_c(i\omega)$ and $E_j(i\omega)$, $j\in \{1,\dots,k\}$, can be validated with~(\ref{eq:mainLMI}).
We will now show how this can be used directly to find subsystem specifications $(\E_1(\omega),\dots,\E_k(\omega))$ as in (\ref{eq:Ej_bound}) for which it is guaranteed that the requirement $\E_c(\omega)$ as in (\ref{eq:Ec_bound}) is satisfied.

Consider the system (\ref{eq:Gc}) and error dynamics (\ref{eq:EcN}) as in Figure~\ref{fig:EcN}. 
Let $\omega \in \R$ and assume that the requirement $\E_c(\omega)$ as in (\ref{eq:Ec_bound}) is given.
Consider the optimization problem
\begin{align}
\label{eq:top_down}
\textrm{given} \quad & V_c(\omega), W_c(\omega) \\
\textrm{minimize} \quad & \tr(V^{-2}(\omega)) + \tr(W^{-2}(\omega)) \nonumber \\
\textrm{subject to} \quad & \left[\begin{array}{cc}
W^{-2}(\omega)D_r^{-1} 	& 	N^H(i\omega) \nonumber \\
N(i\omega)				& 	V^{-2}(\omega)D_\ell
\end{array}\right] \succ 0, \nonumber \\
\quad & V(\omega) \in \mathbf{V}, W(\omega) \in \mathbf{W}, (D_\ell, D_r) \in \mathbf{D}. \nonumber
\end{align}
It follows from Theorem~\ref{thm:main}, that for any feasible solution to (\ref{eq:top_down}), we have that our requirement on the interconnected system $E_c(i\omega) \in \E_c(\omega)$ is satisfied if $E_j(i\omega) \in \E_j(\omega)$ as in (\ref{eq:Ej_bound}) for all $j \in \{1,\dots,k\}$.
%\begin{equation}
%\bar{\sigma}\bigl(V_{c} E_c(i\omega)W_{c}\bigr) < 1.
%\end{equation}
%\end{thm}
%\begin{pf}
%We prove the theorem with the following remarks.
%\begin{enumerate}
%\item For any feasible solution to (\ref{eq:top_down}), condition (\ref{eq:mainLMI}) and subsequently the rest of Theorem~\ref{thm:main} is satisfied.
%\item For $j\in\{1,\dots,k\}$, the following inequalities are equivalent:
%\begin{align}
%\label{eq:proof2}
%\bar{\sigma}(E_j(i\omega)) &\leq \E_j(\omega) = \bar{\sigma}(W_j J_{p_j,m_j} V_j), \nonumber \\
%W_j^{-1}|E_j(i\omega)|V_j^{-1} &\leq  J_{p_j,m_j},  \nonumber \\
%|W_j^{-1}E_j(i\omega)V_j^{-1}| &\leq  J_{p_j,m_j}/\bar{\sigma}(J_{p_j,m_j}).
% \end{align}
%\item Similarly, the following inequalities are equivalent:
% \begin{align}
%\label{eq:proof3}
%|E_c(i\omega)| &\leq \E_c(\omega) = V_c^{-1} J_{p_c,m_c} W_c^{-1}, \nonumber \\
%V_c|E_c(i\omega)|W_c &\leq J_{p_c,m_c},  \nonumber \\
%|V_cE_c(i\omega)W_c| &\leq J_{p_c,m_c}.
% \end{align}
% \item Given $A\in \C^{n\times m}$, if $\bar{\sigma}(A)\leq 1$, then we have $|A| \leq J_{n,m}$. 
%Therefore, if Theorem~\ref{thm:main} is satisfied, then (\ref{eq:proof2}) and (\ref{eq:proof3}) are satisfied, which proves the theorem. \hfill \ \qed
%\end{enumerate}
%\end{pf}

With (\ref{eq:top_down}), an optimization problem to find a solution to the problem as stated in Section~\ref{sec:Problem} is given.
Namely, we can compute local error requirements $(\E_1(\omega),\dots,\E_k(\omega))$ given the global error requirement $\E_c(\omega)$.
Solving the optimization problem, i.e., minimizing $\tr(V^{-2}(\omega)) + \tr(W^{-2}(\omega))$, is relatively trivial by iteratively solving for $V, W$, and $D$, similar to D-K iteration (see \cite{zhou1998}):
\begin{enumerate}
\item Initially, set $d_j = d_c = 1, j=\{1,\dots,k\}$.
\item Relax $V^{-2}(\omega)$ to diagonal matrix $\mathcal{V} := V^{-2}(\omega)$ and $W^{-2}(\omega)$ to diagonal matrix $\mathcal{W} := W^{-2}(\omega)$ and fix $D_r$ and $D_\ell$; the optimization problem (\ref{eq:top_down}) is then linear and $\tr(\mathcal{V}) + \tr(\mathcal{W})$ can be minimized with semi-definite programming (SDP) tools.
\item Fix $V(\omega)$, $W(\omega)$ at the solutions of step (2) and keep $d_c = 1$ fixed. Find the scaling matrices $D_r$ and $D_\ell$ that maximizes $\gamma$ while satisfying the inequality 
\begin{equation}
\label{eq:LMI_solve}
V^{-2}D_\ell - N(i\omega)W^{2}D_rN^H(i\omega) \succ \gamma
\end{equation}
with SDP tools. 
Note that for $\gamma = 0$, the matrix inequality (\ref{eq:LMI_solve}) is equivalent to (\ref{eq:mainLMI}), as proven in the proof of Theorem~\ref{thm:main}. 
By maximizing $\gamma$, the cost function $\tr(V^{-2}(\omega)) + \tr(W^{-2}(\omega))$ can be minimized further in the next iteration.
\item Repeat step (2) and (3) until sufficient convergence in $V(\omega)$, $W(\omega)$ is reached, i.e., $\tr(V^{-2}(\omega)) + \tr(W^{-2}(\omega))$ is no longer decreasing (significantly).
\end{enumerate}
The choice of cost function $\tr(V^{-2}(\omega)) + \tr(W^{-2}(\omega))$ allows to relax the optimization problem (\ref{eq:top_down}) to be easily solved iteratively with SDP solvers.
Additionally, in general, we aim to find a solution in which $V_j(\omega)$ and $W_j(\omega)$ are as ``large'' as possible, which allows for more error in the subsystems, which in turn allows for further reduction of the system as a whole.
Note that within~(\ref{eq:mainLMI}), there is an infinite number of possible combinations $(\E_1(\omega),\dots,\E_k(\omega))$ that guarantee the satisfaction of the requirement $\E_c(\omega)$. 
However, by choosing the cost function $\tr(V^{-2}(\omega)) + \tr(W^{-2}(\omega))$ in~(\ref{eq:top_down}), given $\E_c(\omega)$, the solution converges to a single distribution of subsystem accuracy requirements $(\E_1(\omega),\dots,\E_k(\omega))$.
\begin{rem}
\label{rem:cost_extension}
The advantage of this cost function is that it automatically penalizes individual elements in $V(\omega)$ and $W(\omega)$ that are important for the accuracy of the interconnected system and allows for more error on inputs-to-outputs pairs of the subsystem transfer functions that are less important for the overall accuracy of the interconnected system.
Moreover, if additional knowledge on subsystems is available, e.g., we know that one of the subsystems is more difficult to reduce than the others, the cost function can be trivially extended to 
\begin{align}
\sum_{j=1}^k \alpha_j\left(\tr(V_j^{-2}(\omega)) + \tr(W_j^{-2}(\omega))\right),
\end{align}
where $\alpha_j$ is a weighting variable used to provide some control over the distribution of subsystem requirements in the solution to the optimization problem (\ref{eq:top_down}).
\end{rem}
However, specifying the exact definition of an ``optimal'' distribution of $(\E_1(\omega),\dots,\E_k(\omega))$, and, furthermore, finding this distribution, are still open problems. 
There are various arguments explaining why these problems are not trivial, one of which is the fact that reducing the order of a subsystem generally leads to discrete steps in which the error increases. 
We expect that to find a (sub)optimal distribution of requirements, a heuristic approach, in which communication between subsystems takes place, is required.

In the next section, we will show using an illustrative example from structural dynamics that minimizing the cost function $\tr(V^{-2}(\omega)) + \tr(W^{-2}(\omega))$ is sufficient to compute a distribution of subsystem error requirements that allows for significant reduction of each of the subsystems given, a required $\E_c(\omega)$.

\section{Example}\label{sec:Example}
In this section, we show on a mechanical system consisting of three interconnected beams, as illustrated schematically in Figure~\ref{fig:example}, that the top-down approach can be used to determine frequency-dependent accuracy requirements $\E_1(\omega),\E_2(\omega),\E_3(\omega)$ for the reduced-order models of the three beams based on given requirements $\E_c(\omega)$ for the accuracy of the reduced-order, interconnected model, allowing for the independent reduction of these subsystems.

Subsystems 1 and 3 are cantilever beams which are connected at their free ends to free-free beam 2 with translational and rotational springs.
The stiffness of both translational interconnecting springs is $k_1^t = k_2^t = 1 \times 10^5$ N/m.
The stiffness of both rotational interconnecting springs is $k_1^r = k_2^r = 1 \times 10^3$ Nm/rad.
The external input force $u_c$ [N] is applied to the middle of subsystem 2 in the transversal direction.
The external output displacement $y_c$ [m] is measured at the middle of subsystem 3 in the transversal direction.

\begin{figure}
   \centering
   \includegraphics[scale=0.8]{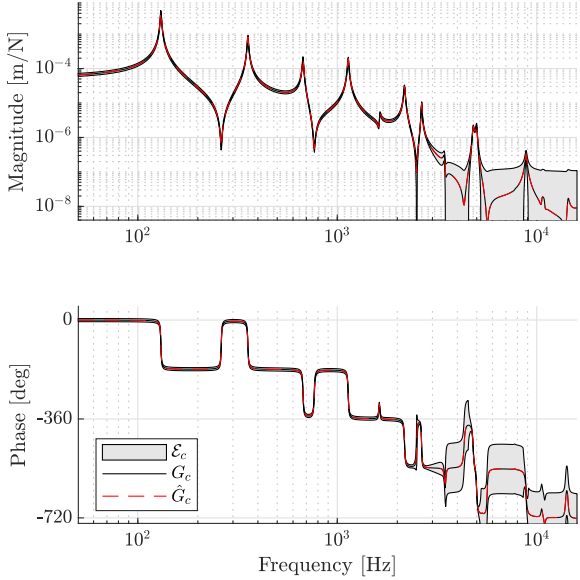} 
   \caption{The Bode plots of the high-order interconnected model $G_c$ with $n_1 = 200$, $n_2 = 84$ and $n_3 = 120$, the areas in which a system satisfies error requirements $E_c(i\omega) \in \E_c(\omega)$, and of the computed reduced-order interconnected model $\hat{G}_c$ with $r_1 = 19$, $n_2 = 17$ and $n_3 = 15$.}
   \label{fig:TD_Gc}
\end{figure}

Each beam/subsystem is discretized by linear two-node Euler beam elements (only bending, no shear, see~\cite{craig2006}) of equal length. 
Per node we have one translational degree of freedom (dof), i.e., a transversal displacement, and one rotational dof.
For each beam, viscous damping is modelled using 1\% modal damping.
%Note that we use the same example system as in \cite{janssen2022_automatica}, with slightly altered parameter values, to show that the method can also be applied to systems with weak damping.
Physical and geometrical parameter values of the three beams and information about finite element discretization, the number of states, and the number of subsystem inputs and outputs are given in Table~\ref{tab:parameters}.
The Bode plot of the unreduced system $G_c$ is given by the black line  inFigure~\ref{fig:TD_Gc}.

\begin{figure}
   \centering
   \includegraphics[scale=0.8]{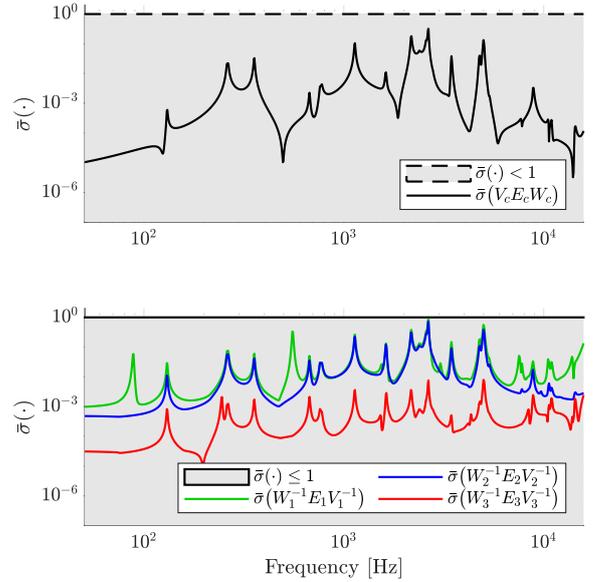} 
   \caption{Largest singular value plots, indicating that the reduced-order subsystem models satisfy the required $E_j(i\omega) \in \E_j(\omega)$ for $j=1,2,3$ and, consequently, that the reduced-order interconnected model satisfies $E_c(i\omega) \in \E_c(\omega)$.}
   \label{fig:TD_all}
\end{figure}

For this system, the top-down approach is applied with the following steps:
\begin{figure}
   \centering
   \includegraphics[scale=1, page=5]{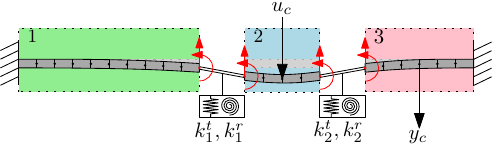} 
   \caption{Example system: Two cantilever beams (Subsystems 1 and 3) connected on their free ends to free-free beam (Subsystem 2) with translational and rotational springs.}
   \label{fig:example}
\end{figure}
\begin{table}[]
\caption{Parameter values of each subsystem.}
\label{tab:parameters}
\begin{tabular}{l|lll}
Parameter                             & Subsys. 1       & Subsys. 2       & Subsys. 3       \\
\hline                        
Cross-sect. area [m$^2$]       & $1\times 10^{\text{-}6}$ & $1\times 10^{\text{-}6}$ & $1\times 10^{\text{-}6}$ \\
2nd area moment [m$^4$]        & $1\times 10^{\text{-}9}$ & $1\times 10^{\text{-}9}$ & $1\times 10^{\text{-}9}$ \\
Young's modulus [Pa]           & $2\times 10^{11}$        & $2\times 10^{11}$        & $2\times 10^{11}$ \\
Mass density [kg/m$^3$]        & $8\times 10^{3}$         & $8\times 10^{3}$         & $8\times 10^{3}$  \\
Modal damping [-]              & $0.01$                   & $0.01$                   & $0.01$  \\
Length [m]                     & $1$                      & $0.4$                    & $0.6$             \\
\# of elements [-]             & $50$                     & $20$                     & $30$              \\
\# of inputs $m_j$ [-]         & $2$                      & $5$                      & $2$       \\    
\# of outputs $p_j$ [-]        & $2$                      & $4$                      & $3$        \\     
\# of states $n_j$ [-]         & $200$             	      & $84$                     & $120$       \\                  
%\# of states $r_j$ [-]         & $19$             	      & $17$                     & $15$       \\   
\end{tabular}
\end{table}
\begin{enumerate}
\item All frequencies $\omega$ over a grid of 1000 logarithmically equally spaced points in the interval $[10^{2.5},10^5]$ rad/s are evaluated.
For these frequencies, a frequency-dependent accuracy requirement $\E_c(\omega)$ as in (\ref{eq:Ec_bound}) is provided by the user. 
In this example, $V_c^{-1}(\omega)$ is given as some fraction $\beta_1$ of $|G_c(i\omega)|$, which is bounded below by $\beta_2$,  
\begin{equation}
V_c^{-1}(\omega) = \max\{\beta_1 |G_c(i\omega)|, \beta_2\},
\end{equation} 
where $\beta_1=0.1$ and $\beta_2=5\times 10^{-7}$ m/N and $W_c(\omega) = 1$.
The resulting accuracy requirement is indicated by the grey areas in Figure~\ref{fig:TD_Gc}.
Any error $E_c(i\omega)$ satisfies $E_c(i\omega)\in\E_c(\omega)$ for the given frequencies $\omega$ if and only if $\bar{\sigma}(W_c(\omega)E_c(i\omega)V_c(\omega)) < 1$, i.e., is entirely in the grey area in the top figure of Figure~\ref{fig:TD_all}.
\item The optimization problem (\ref{eq:top_down}) is solved, which results in subsystem requirements $(\E_1(\omega),\E_2(\omega),\E_3(\omega))$, for the given frequencies $\omega$. 
These requirements consist, for each subsystem $j = 1,2,3$, of diagonal scaling matrices $W_j(\omega)$ and $V_j(\omega)$ that describe the scaling of individual input-output pairs in the requirements. 
Any error $E_j(i\omega)$ satisfies $E_j(i\omega)\in\E_j(\omega)$ for the given frequencies $\omega$ if and only if $\bar{\sigma}(W_j^{-1}(\omega)E_c(i\omega)V_j^{-1}(\omega))\leq 1$, i.e., is entirely in the grey area in the bottom figure of Figure~\ref{fig:TD_all}.
\end{enumerate} 
With the proposed approach, in principle, any MOR method can be used for which it is possible to find a reduced-order subsystem such that the computed requirements are satisfied, i.e., $E_j(i\omega) \in \E_j(\omega)$.
It is even possible to use different MOR methods for each subsystem.
However, as the purpose of this work is to show how subsystem error requirements can be determined from the top down, we apply a standard MOR method to all of the subsystems.
\begin{enumerate}
\setcounter{enumi}{2}
\item Using the computed $(\E_1(\omega),\E_2(\omega),\E_3(\omega))$, model reduction techniques can be used to construct reduced-order subsystems satisfying $E_j(i\omega) \in \E_j(\omega), j = 1,2,3$, for all frequencies $\omega$. 
In this example, we use frequency-weighted balanced truncation (FWBT) \citep{enns1984} to reduce the individual subsystems. 
In FWBT, we can minimize $\|\hat{W}_j(G_j-\hat{G}_j)\hat{V}_j \|_\infty$, where $\hat{V}_j$ and $\hat{W}_j$ are transfer function estimates fitted with a minimum-phase transfer function \citep[Chapter 6.5]{boyd2004}, of the computed weighting functions $V_j$ and $W_j$, respectively. 
See \cite{gugercin2004} for more details on FWBT. 
In Figure~\ref{fig:TD_all}, for each subsystem, we show that $E_j(i\omega) \in \E_j(\omega)$ with the green, blue and red lines, respectively.
With FWBT, the reduced-order models can be reduced to $r_1 = 19$, $r_2 = 17$ and $r_3 = 15$ while satisfying the given subsystem error requirements.
\item To validate the approach, we show that $E_c(i\omega) \in \E_c(\omega)$ is indeed satisfied for the interconnected system, as indicated by the black line in Figure~\ref{fig:TD_all}.
Additionally, we show the reduced-order interconnected system in the top figure in Figure~\ref{fig:TD_Gc} with a dashed red line, and see that the reduced-order system indeed satisfies the requirement.
\end{enumerate} 
With these steps, we show that, for this system, it is possible to determine error requirements at a subsystem level that 1) guarantee that the overall requirements on the accuracy of the reduced-order model for the interconnected system are satisfied and 2) allow for enough ``room'' for the significant reduction of the subsystem models.
Even with the standard MOR technique we apply on a subsystem level, i.e., FWBT, it is already possible to reduce the number of states of the interconnected system from $\sum n_j = 404$ to $\sum r_j = 51$ within the given requirements.
If more involved MOR methods are applied, the subsystem models can potentially be reduced even further within the computed accuracy specifications $\E_j(\omega)$.

\section{Conclusions}\label{sec:Conclusion}
%Subsystems are increasingly developed, modeled, and analyzed independently by distinct teams. 
%With modular model reduction, the complexity of a system consisting of multiple interacting subsystems can be reduced at a subsystem level by the individual teams.
%If design changes are made to a single subsystem, only the ROM of that subsystem needs to be updated.
%Additionally, the computationally challenging reduction of one high-dimensional model is avoided and the interconnection structure of the original high-order system is preserved.
%
%Reducing the complexity of subsystems does not only decrease the accuracy of the subsystem itself, but also of the overall interconnected system.
In this paper, we demonstrate how, for models of interconnected LTI subsystems, accuracy requirements on the interconnected system can be translated to accuracy requirements of subsystems.
With these requirements, modular model reduction can be applied while guaranteeing the required accuracy of the overall interconnected system.

The approach is based on the reformulation of subsystem reduction errors to weighted uncertainties.
This allows for mathematical tools from the field of robust performance analysis to be applied.
We show that with this reformulation, a single matrix inequality can be used to analyze if accuracy requirements at the subsystem can guarantee that given accuracy requirements at the interconnected system level are satisfied.
Moreover, we propose an optimization problem that can be used to compute these subsystem accuracy requirements and we show how this problem is solved.
Finally, the approach is illustrated with a structural dynamics example, for which the complexity in terms of the number of states in the overall system can be reduced by at least 87\% for the given requirements.

%Future work consists of applying the method to the model of an industrial system.
%Furthermore, the optimization problem can be extended such that more details on the reduction of subsystems can be taken into account, for example by including the extension proposed in Remark~\ref{rem:cost_extension}.
%As another example, if the reduction of one of the subsystems that does not fully ``utilize'' the allowed error requirement, the requirements on the other subsystems can be relaxed accordingly.
%Such heuristics approach can be used to decrease the complexity of the interconnected system even further.

\begin{ack}
This publication is part of the project Digital Twin with project number P18-03 of the research programme Perspectief which is (mainly) financed by the Dutch Research Council (NWO).
\end{ack}

\bibliography{ifacconf}             % bib file to produce the bibliography

\end{document}